# A Review of HPC-Accelerated CFD in National Security and Defense

Author: James Afful*

Email: affulj@iastate.edu

Affiliation: Independent Researcher, Ames, Iowa, USA.

## Abstract

Using High-Performance Computing (HPC), Computational Fluid Dynamics (CFD) now serves as an essential component in defense-related national security applications including missile interception and hypersonic propulsion as well as naval stealth optimization and urban hazard dispersion. This review combines two decades of open-source and public-domain research on HPC-accelerated CFD in defense, addressing three key questions: Which security-sensitive simulations have utilized open-source CFD frameworks such as OpenFOAM, SU2 and ADflow? Which HPC techniques, such as MPI domain decomposition and GPU acceleration together with hybrid parallelism best enhance open-source frameworks to manage large defense CFD simulations? Which technological advancements and research voids currently drive the directional development of the field? Examining several research studies sourced from NASA, DoD HPC centers, and academic institutions, scientific contributions have been classified into air, maritime, and space domains. Modular frameworks like NavyFOAM and SU2 and ADflow's adjoint-based solvers show how custom open-source solutions support workflows with rapid completion of multi-million cell simulations. The conclusion highlights new trends that combine exascale readiness with machine learning surrogate models for real-time CFD applications and interdisciplinary HPC-driven multi-physics integration to deliver practical insights for improving CFD use in defense research and development.

# INTRODUCTION

## 1. Context and Rationale

### 1.1 Evolution of High-Performance Computing (HPC)

The development of High-Performance Computing (HPC) has shown remarkable progress since its inception in the mid-20th century. The earliest definition of "supercomputing" described specialized mainframes which executed hundreds of thousands of operations per second and were regarded as state-of-the-art during that period [27]. The development of transistor technologies led to the creation of vector-based architectures which enabled large scientific computations to achieve unprecedented speeds compared to traditional systems. During the 1980s and 1990s government investments in national laboratories and defense agencies enabled massively parallel processing (MPP) development while solving computational problems in fields such as aerodynamics and nuclear physics [11, 38]. The development of parallel cluster computing allowed HPC systems to connect commodity hardware through high-speed networks to expand the available compute cores for simultaneous numerical task execution [14]. The HPC sector advanced further during the 21st century because multi-core processors alongside general-purpose graphics processing units (GPGPUs) and additional accelerators formed essential parts of expansive supercomputing platforms [27]. We have moved beyond petascale computing that reached $10^{15}$ FLOPS to enter the exascale computing domain, which now achieves at least $10^{18}$ FLOPS [61, 86]. The rapid increase in computational power has continuously transformed how scientists and engineers develop real-world phenomenon models [60]. Defense agencies around the world have maintained a leading position in HPC innovation by supporting research into hardware and numerical algorithms, which enables them to conduct complex military simulations. Defense-related HPC simulations have historically tackled nuclear stockpile stewardship and ballistic missile defense scenarios while also cracking cryptographic codes and performing large-scale aerodynamics and fluid flow analyses, which hold special significance for this review [18, 31, 32].

Computational Fluid Dynamics (CFD) serves as an essential tool for defense applications by facilitating simulations of fluid flows around complex shapes. CFD combines numerical analysis and PDE solvers with computing architecture to model and forecast fluid movements around intricate shapes [34, 39, 79, 82]. CFD methods prove crucial for analyzing aerodynamic and aero-thermodynamic phenomena which support various defense applications related to national security. CFD simulations that deliver both efficiency and precision help determine the best designs for missiles as well as their flight paths and control surface performance [18]. Precise modeling of lateral jet interactions in systems such as the Terminal High Altitude Area Defense (THAAD) interceptor affects ballistic missile defense performance outcomes [9, 8]. Modern fighter jets and bombers, along with reconnaissance vehicles, depend on high-performance computing CFD simulations to minimize radar cross-sections through aerodynamic designs and to understand complex airflow patterns around dynamic control surfaces [21]. Engineers who can accurately forecast high-angle-of-attack dynamics and fluid-structure interactions achieve lower flight-testing expenses and improved flight safety [24]. Submarines and surface ships undergo large-scale free-surface simulations to enhance hull design performance in areas of speed and stealth capabilities as well as maneuverability. Naval defense systems benefit from NavyFOAM's HPC capabilities, which enable accurate modeling of critical multiphase flows and wave interactions according to [49]. Computational fluid dynamics extends beyond aerodynamic uses to become crucial for simulating the spread of chemical, biological, and radiological substances in urban settings and battlefields [50, 64, 72]. Rapid urban flow simulation capabilities enable effective planning for evacuation routes and sensor placement while improving first-responder preparedness. High-performance computing-based computational fluid dynamics serves as a critical tool for modeling chemically reactive high-temperature flows in advanced propulsion and hypersonic vehicle research where traditional fluid models fail [28]. Research applications span next-generation scramjet development and high Mach number anti-ballistic missile systems.

HPC-accelerated CFD functions as a force multiplier in defense applications by making previously impractical complex fluid simulations achievable. Traditional wind tunnel testing retains relevance but faces prohibitive costs and constraints related to size and speed limitations. High-fidelity CFD acts both as a supplementary and occasional substitute approach to speed up design iteration and enhance mission success rates through prototype refinement [75]. High-performance computing (HPC) acceleration has become crucial for CFD workflows due to its ability to handle the complex requirements of modern defense applications.

CFD technology originated before the HPC revolution, but today's defense-focused analysis needs billions of grid cells together with advanced turbulence modeling approaches such as Detached-Eddy Simulation or Large-Eddy Simulation and Multiphysics coupling to effectively simulate combustion processes and fluid-structure interactions as well as shock waves [17, 54]. The computational requirements increase alongside the growing complexity of these simulations. HPC achieves multi-level parallelism through distributed memory (MPI), shared memory (OpenMP), and GPU accelerators (CUDA, OpenACC) to process massive CFD problem sizes [55, 63]. Effective load balancing during missile aerodynamics and multi-vehicle interaction simulations requires hybrid parallel approaches. In the past, a single iteration required days to weeks to reach convergence using traditional CFD methods, but modern HPC architectures, alongside GPU acceleration significantly reduce simulation runtimes to mere hours or minutes which makes real-time and near real-time simulations feasible [4, 27, 51]. Shorter turnaround times in defense result in accelerated design cycles and agile mission planning while enabling broader design space exploration within limited timeframes. High-performance computing facilitates the detailed analysis of fine-scale turbulence structures along with transient shock reflections and boundary-layer phenomena that are essential for stealth design and supersonic engine inlets according to [61]. Achieving high fidelity enables a strategic edge by uncovering subtle aerodynamic effects that lower-resolution simulations or simpler models often overlook. High-performance computing systems enable extensive exploration of design spaces through multiple parametric variations. HPC-based CFD simulations combined with uncertainty quantification methods or robust optimization frameworks enhance system reliability and performance when dealing with uncertain or extreme operating conditions.

### 1.2 Open-Source Research Contributions

The development of defense-oriented CFD technologies typically occurs in secrecy because of security classification requirements and proprietary vendor solutions [56]. During the last ten years open-source CFD platforms like OpenFOAM and SU2 have become popular across academic circles and business sectors as well as within some defense-related initiatives [3, 17, 54]. Community-driven innovation along with cost-effectiveness and transparency and ease of collaboration have enabled these outcomes according to [85]. Open-source communities enable swift development and testing of new models whether they involve turbulence closures or advanced boundary conditions. The collaboration of researchers across the globe enables faster code improvements compared to closed-source products. The NavyFOAM project modifies OpenFOAM to serve naval hydrodynamics needs which shows open-source solutions can tackle defense-specific applications [49]. Financial constraints along with requirements for scientific methods that produce reproducible and verifiable results prompt government agencies and defense industry players to embrace or modify open-source solutions. Building reliable models requires the capability to examine and adjust source code because mission-critical design choices depend on each code line. Open-source CFD enables universities and small defense firms to educate engineers and scientists without the expense of commercial software licenses. The defense talent pool expands to include professionals who understand HPC systems, modern software development processes, and fluid dynamic modeling techniques [7]. The classification status of top-tier HPC and CFD projects does not prevent the publication of core algorithms and HPC scaling strategies along with basic flow physics insights [57]. The scientific community benefits from HPC-based project overviews that provide substantial information about numerical methods and optimization techniques despite any redaction or access restrictions.

# 2. Scope and Objectives of the Review

## 2.1 Research Questions and Objectives

This systematic review investigates HPC-accelerated CFD applications in national security and defense and focuses on examining contributions from open-source frameworks. The research questions that follow were established to ensure structure and cohesion.

1. **RQ1:** *In what ways have open-source CFD frameworks like OpenFOAM, SU2, and ADflow been utilized for solving defense-oriented challenges?*

2. **RQ2:** *Which leading HPC methods including distributed memory parallelism, GPU acceleration, and hybrid MPI/OpenACC are used to scale open-source CFD programs for large defense simulations?*
3. **RQ3:** *What emergent trends or gaps does the literature reveal, and how might these inform future developments of HPC-accelerated CFD in defense contexts?*

The following parts of the review will describe the Methodology which includes the literature collection process and the criteria for inclusion and exclusion. The Findings and Discussion section presents the review's results which have been organized under themes including HPC scaling strategies along with open-source code capabilities and defense application subdomains. The Research Gaps and Future Directions section identifies critical deficiencies in HPC-based CFD for defense applications and suggests future solutions, such as exascale preparedness and machine learning implementation alongside real-time HPC for essential fluid dynamics operations and security measures. The Conclusions and Recommendations sections outline crucial takeaways which are intended for defense agencies and HPC centers as well as the open-source community. The structure of the document provides clarity by leading readers from the introduction's goals to an in-depth analysis of existing research literature.

# 3. Methodology

## 3.1 Search Strategy

The choice of databases and search engines is an important step in any systematic review. The initial goal was to capture a broad swath of published studies— ranging from peer-reviewed journal papers to conference proceedings, technical reports, and thesis/dissertation work. To this end, multiple resources were targeted: Google Scholar, chosen for its breadth in indexing not only journal articles but also conference papers, theses, white papers, and various technical documents; IEEE Xplore and arXiv, highly relevant for HPC and CFD research because it indexes key conferences (such as the IEEE International Conference on High Performance Computing) and journals covering computational science, fluid mechanics, and defense-oriented research; and ResearchGate, which is useful for discovering preprints, technical updates, and author-submitted copies of works that may not appear in other databases, as well as for tracking authors known for HPC-accelerated CFD in defense contexts. We also utilized the NASA Technical Reports Server (NTRS) and STI Repository, given NASA's longstanding role as a powerhouse in HPC-driven CFD with outputs and collaborations relevant to national defense. By drawing on these multiple sources, selection bias was reduced, and the search was not confined to a narrow range of publications or only to high-impact journals— an approach considered a best practice in systematic reviews that span multidisciplinary or defense-related topics [81].

In this project, query sets were refined iteratively during an initial pilot period of screening titles and abstracts. The final set of search strings included logical operators (AND, OR, " " for exact phrases) and truncation where needed, alongside parentheses to group concepts; for example, combinations such as "HPC CFD" OR "High-Performance Computing" AND "Computational Fluid Dynamics," "Defense" OR "National Security" AND "HPC" OR "High-Performance Computing," "Open-Source CFD" AND "HPC" OR "GPU acceleration" OR "Parallel computing," "Defense" AND "HPC CFD" AND "Open-Source," and "CFD" AND "HPC" AND ("Missile" OR "Interceptor" OR "Aircraft" OR "Naval" OR "Urban security" OR "CBR defense") were used. In addition to these primary strings, synonyms and variations— such as "aerospace CFD," "military CFD," and "missile aerodynamics simulation" were used to capture works that might not explicitly label themselves as "CFD in defense." We also recognized that "high-performance computing" can appear under multiple terms (HPC, supercomputing, parallel computing, exascale computing), so we used each to broaden the net. Furthermore, in aiming to capture open-source aspects, we referenced known tools like "OpenFOAM," "SU2," "ADflow," "solids4Foam," and "NavyFOAM," as well as HPC frameworks that are often open or community-developed (e.g., CHSSI and HPC modernization efforts). This comprehensive approach minimized the likelihood of missing relevant studies that did not explicitly bill themselves as "open-source CFD" but might still have relied on or extended open-source solvers.

To capture the contemporary state while still acknowledging significant historical contributions to HPC-based CFD in defense, we focused our primary search on studies published between 2005 and 2025, with the latter date including in-press and forthcoming works that might only appear in preprint form. However, we allowed select exceptions for seminal pre-2005 papers— such as foundational articles (e.g., Boris, 1989) [11] or older NASA/Air Force HPC conference proceedings— that are frequently cited as the basis of modern HPC CFD and thus provide critical historical context. Similarly, because HPC technology evolves rapidly, we extended our scope to include recent preprints or accepted manuscripts dated through 2025, paying particular attention to materials discussing exascale readiness.

## 3.2 Screening Process

In the Identification phase, broad queries using the search strings noted in the preceding section were conducted, capturing all potentially relevant titles and abstracts from each database. During this step, each search date, the keywords used, and the database or platform accessed were documented. Next, in the Screening (Title/Abstract) Level, potentially relevant records then advanced to the next phase, while items that were clearly outside the domain— such as HPC for purely biomedical fluid or cryptographic purposes with no connection to CFD— were discarded. Subsequently, during the Eligibility (Full-Text) Review, the complete text for the papers that had passed the initial screening were then accessed. This step validated the presence of HPC-accelerated CFD for defense or security applications and confirmed an open-source dimension. Once the final set of papers was determined, a structured data extraction sheet was used to systematically record key information. Each record in the final corpus was evaluated across several dimensions, starting with bibliographic details (title, authors, year of publication, publication type, and affiliations), followed by HPC features— such as the parallelization strategy (MPI, OpenMP, hybrid CPU-GPU, exascale approaches) and specific HPC architectures used (for instance, the size of the cluster or the HPC center involved). We then noted the CFD framework and version, documenting whether the code base was purely open-source or included licensed modules, along with any version numbers or repository links the authors provided. Next, we categorized the type of defense application— covering missile or interceptor aerodynamics, naval or ship hydrodynamics, aircraft (fighters, bombers, UAVs) or stealth considerations, chemical and biological defense, and urban fluid dynamics for homeland security— and recorded the scale of each simulation, including the number of mesh cells, HPC node count, typical runtime, or speedup achieved. Performance benchmarks or validation efforts examined, noting any reported runtimes, parallel scalability metrics, or code performance comparisons across different architectures, and whether real experimental data were used for validation. Finally, there was a check for open-source contributions, such as whether the solver or a portion of it was publicly available; whether input datasets or scripts were provided, and if a GitHub/GitLab repository or institutional open-access repository was referenced. This approach ensured uniform processing of each paper.

# 4. HPC and Its Role in Defense & National Security

The Exascale Computing Project (ECP), along with leading-edge systems such as Frontier at Oak Ridge National Laboratory, focus on defense-critical applications, including nuclear stockpile stewardship and large-scale aerodynamic design in the United States [33, 62]. Exascale efficiency relies on software scalability across millions of cores, which HPC code modernization efforts alongside advanced numerical algorithms and machine learning surrogates together with domain-specific frameworks such as NASA's CFD Vision 2030 are currently working to solve [16, 75].

## 4.1 Trends in Parallelization Paradigms (MPI, OpenMP, CUDA, Etc.)

The evolution of software paradigms for parallel computing has accompanied the scaling of HPC hardware to meet emerging demands which now span distributed-memory models, shared-memory approaches and frameworks specific to accelerators [36]. Distributed-memory, with the Message Passing

Interface (MPI) at its core, remains the de facto standard: MPI serves as the coordinating framework for fine-grained interaction among discrete processes in large clusters that consist of thousands of nodes [53]. The NPARC Alliance Flow Simulation System which is focused on defense applications uses MPI to divide massive flow fields into smaller subdomains for parallel solution processing while exchanging boundary information at each iteration step. Shared-memory models create parallel processing capabilities at the node level. HPC nodes today often feature multiple CPU sockets with many cores each which makes OpenMP a standard choice for threading loops and tasks in shared-memory environments [10]. Programmers access GPU acceleration capabilities by using either CUDA for NVIDIA devices or OpenCL for vendor-neutral applications to modify their performance-sensitive kernels to execute across thousands of GPU threads [27, 29]. Ballistic missile flow studies benefit from GPU-driven methods that dramatically reduce simulation times and provide essential response speed during intense defense operations [18]. OpenACC serves as a higher-level pragma-based approach that streamlines GPU offloading by eliminating the need for comprehensive CUDA code rewrites according to [55]. The integration of MPI across nodes with CUDA or OpenACC and OpenMP within nodes has become standard practice in high-performance computing-based computational fluid dynamics. Defense applications require multi-level parallelism to run complex simulations including both urban-scale agent dispersion and high-fidelity supersonic flow scenarios [40, 72]. The movement toward exascale computing introduces new complexities such as resilience requirements and fault tolerance which must be managed at previously unseen levels of concurrency [75]. CFDNet [63] alongside data-driven adaptivity frameworks reduce communication overheads in large-scale CFD simulations and exascale systems need sophisticated memory management coupled with hierarchical parallelism and domain-specific languages. Despite the exascale projects being at early development stages as prototypes they demonstrate that defense HPC initiatives will need stronger partnerships between domain scientists numerical analysts and system architects.

## 4.2 Discussion of Exascale Initiatives and Next-Generation HPC Architectures

The Department of Energy (DOE) identified exascale computing as a fundamental shift in massive simulation methods beyond just hardware improvements while defense agencies around the globe consider it a key strategic objective. The Exascale Computing Project (ECP) in the United States brings together national labs, academic institutions, and industry partners to advance the development of exascale-ready hardware and software alongside application capabilities. Numerous ECP applications which impact national security through nuclear weapons stewardship and hypersonic missile aerodynamic design demand codes that can scale to hundreds of thousands of GPUs while handling faults and providing distributed parallel I/O software libraries [51]. The research efforts of NASA's CFD Vision 2030 [16, 75] aligns with the Department of Defense's HPC modernization projects through shared ECP deliverables which connect HPC research foundations to defense operational requirements. The main interest in this case, lies in U.S. defense applications, but comparable exascale infrastructure investments exist globally through projects such as the European Union's EuroHPC program and China's Sunway TaihuLight exascale prototype along with Japan's RIKEN Fugaku system. Advanced fluid simulations for scramjet technologies and high-speed rail aerodynamics utilize these platforms for dual-use or defense-focused applications. International HPC competitions drive advancements across data movement techniques along with memory hierarchy designs and specialized accelerator technologies which are essential for large-scale fluid mechanics applications. In addition to raw FLOP counts, exascale systems must focus on energy efficiency since power requirements can exceed tens of megawatts [49, 78] and need to address memory/storage because HPC nodes now incorporate high-bandwidth memory (HBM) alongside parallel file systems to manage massive data volumes in defense applications [62]. To scale effectively across millions of cores experts need sophisticated partitioners alongside advanced solvers and dynamic load balancing techniques which address localized flow phenomena including shock-vortex interactions in hypersonic conditions [66]. The expansion of AI integration continues as hybrid high-performance computing with artificial intelligence systems use large data sets to train surrogate models which accelerate full-scale computational fluid dynamics and provide real-time defense analytics capabilities [11, 63]. Defense laboratories benefit from exascale HPC because these platforms enable comprehensive, unsteady multiphysics simulations for ballistic missile modeling and stealth aircraft design that reduce both development expenses and testing timeframes [25].

# 5. Importance of HPC in National Security

## 5.1 Why HPC Is Critical for Advanced Simulations

Advanced simulations depend critically on high-performance computing (HPC) to address modern defense challenges. Defense systems today require HPC-based CFD to address complex multiphysics challenges like intercepting hypersonic projectiles and naval fleet optimization for reduced radar and thermal signatures to manage phenomena such as multi-phase reactive flows for missile defense and naval environment interactions including free-surface hydrodynamics and structural vibrations [9, 49, 64]. HPC enables near real-time decision-making capabilities which are demonstrated in urban defense models integrating meteorological and sensor data in real time [64, 72]. Simulation updates occur quickly to provide guidance for both emergency responders and interceptors yet maintaining strict real-time operations presents challenges without continuous optimizations and support from machine learning replacements [6, 63]. High-fidelity multiphysics modeling which includes fluid-structure interactions, electro-magneto-hydrodynamics, combustion, and chemical kinetics needs HPC capabilities to address complex systems of coupled partial differential equations whether solved by partitioned schemes or monolithic approaches therefore rendering HPC essential for both iterative design processes and urgent defense problem-solving [48, 54].

## 5.2 Real-World Exemplars in Defense

Open-source research and partial disclosures show HPC transforming real-world defense applications even with existing classification restrictions. Ballistic missile defense employs HPC-based CFD techniques to enhance interceptor capabilities by examining side thruster interactions with external flow fields demonstrated in the THAAD system research [18]. Computational high-performance simulations provide critical information for reentry vehicle design through the examination of ablation, ionization, and shock-layer chemistry which supports the optimization between drag reduction and thermal load management [31]. High-performance computing (HPC) enables stealth aircraft designers to minimize radar signatures and infrared emissions while controlling noise through precise aerodynamic shaping and boundary-layer management techniques. By integrating compressible Navier–Stokes equations with structural dynamics or engine performance models through detailed simulations engineers can identify the best inlet geometries and nozzle arrangements that maintain stealth capabilities while preserving flight performance. High-performance computing capabilities enable turbulence analysis at extreme angles of attack which provides essential information about vortical flows across stealth fuselage surfaces or wing-leading edges thereby supporting the development of advanced stealth aircraft designs that combine maneuverability with low visibility [7].

Similar to its role in other engineering domains, HPC supports naval hydrodynamics with tasks including hull form optimization and propeller and rudder analysis [49]. Partitioning free-surface domains and managing millions of calculations per time step requires systems with large memory capacity and high-bandwidth interconnects [51]. HPC-accelerated codes support nuclear weapons management which requires modeling fluid dynamics, radiation transport, and plasma physics at large scales to simulate nuclear explosions instead of conducting live tests [38]. The need for more precise simulation of shock propagation and nuclear reactions has pushed High-Performance Computing (HPC) development from terascale to petascale and then to exascale systems despite many simulation frameworks remaining classified or partially open [11, 14].

High-Performance Computing (HPC) impacts defense sectors not only through simulations but also by fostering innovation through accelerated R&D processes and interdisciplinary technique exchange between fluid dynamics and other fields as well as providing advanced training for computational scientists in national security initiatives [23]. High Performance Computing allows virtual prototyping which decreases expensive physical tests and speeds up design-test iteration cycles while fostering integrated approaches combining structural, thermal, and electromagnetic simulations [1]. Government laboratories and academic institutions work closely with private contractors in this ecosystem while public sharing of partial HPC code improvements stimulates broad advancements in numerical methods and parallel performance. The combined capabilities of HPC technology with advanced materials and data analytics alongside domain-specific expertise are creating a new "digital engineering" phase in defense that enables increasingly complex fluid dynamic simulations for advanced weaponry and protective systems along with intelligence applications as HPC resources keep expanding.

# 6. Survey of Open-Source Literature: HPC-Accelerated CFD in Defense

This section provides a systematic review that examines how open-source or partially open High-Performance Computing (HPC) projects find applications in defense Computational Fluid Dynamics (CFD). This review findings center around the application of open-source and partially open High-Performance Computing (HPC) projects within defense-related Computational Fluid Dynamics (CFD). The literature is categorized according to primary defense areas including Airborne Platforms, Naval and Submarine Hydrodynamics, Space and Missile Defense, Directed Energy and Other Emerging Technologies, and Security & Intelligence Applications. Our analysis brings together principal HPC methods including domain decomposition and GPU acceleration across these subsections), open-source frameworks utilized (OpenFOAM, SU2, etc), turbulence models (RANS, LES, etc. We aim to demonstrate the extensive scope and detailed study of HPC-accelerated CFD research within defense literature that is publicly available while recognizing that classification restrictions can prevent complete access to certain datasets and algorithms.

## 6.1 Airborne Platforms

The design and performance assessment of airborne defense systems such as high-lift aircraft and UAVs through to supersonic or hypersonic missiles necessitate advanced CFD approaches [24, 31], which must include precise turbulence modeling and control surface replication while managing shock structures or chemical reactions in sophisticated propulsion systems. Studies on high-lift UAV aerodynamics and jet performance demonstrate the essential role of open-source high-performance computing accelerated CFD in accurately simulating multi-physics phenomena [7, 21, 47]. The use of HPC-based turbulence models like Detached-Eddy Simulation and hybrid RANS-LES facilitates understanding of vortical flows around stealth configurations despite the limited access to stealth aircraft research according to [24]. The research on supersonic and hypersonic missile flows features foundational work in scramjet technology and real-gas modeling [31] as well as open-source computational fluid dynamics tools like ADflow which support high-speed missile design through adjoint-based optimization and domain decomposition methods. The intricate specifics of ballistic or midcourse intercept situations stay mostly secret while open-source code references exist but vital geometric details and flight parameters are frequently omitted for security purposes [18].

### 6.1.1 Representative Table: Airborne Platforms in Open-Source HPC

A textual summary table presents key references alongside HPC methods, frameworks, mesh scales, turbulence models and findings relevant to the airborne platform domain:

| Reference | HPC Approach | Open-Source Framework | Computing Resources | Mesh Scale | Turbulence Model | Key Findings |
|---|---|---|---|---|---|---|
| [7] | MPI domain decomposition | OpenFOAM | Not specified | 0.01–0.8 million cells | RANS (SST) | Verified OpenFOAM for high-lift flows; HPC gave near-linear scaling |
| [21] | Parallel unsteady solvers | Partially disclosed (Cobalt) | 128 - **1,000** CPU cores | Over 30 million cells (dynamic) | RANS/URANS, DES | Large-scale dynamic stability simulations of control surfaces; HPC essential for real-time control loops. |
| [31] | MPI-based parallel (generic HPC) | GPACT | Not specified | Not specified | Various | HPC for advanced propulsion; highlights complex chemical kinetics in scramjets. |
| [54] | MPI domain decomposition + adjoint | ADflow | **1000**s of cores | Millions of cells depending on design | RANS, some LES modules | Adjoint-based HPC for aerodynamic optimization, easily extended to supersonic regime. |
| [18] | Not specified | Not specified | Defense HPC facilities | Millions of cells for jet interactions | Not specified | Modeled lateral jet interaction on THAAD interceptor; HPC reduced design iteration times significantly. |

## 6.2 Naval and Submarine Hydrodynamics

Naval hydrodynamics involves complex phenomena including free-surface flows and propeller-rudder interactions as well as wave-structure effects and multiphase conditions like cavitation which requires high-fidelity HPC-based CFD to achieve stealth and efficiency [22, 49]. Academic institutions and government-sponsored research projects use open-source software to analyze complex unsteady free-surface domains by utilizing HPC clusters with hundreds to thousands of processing cores. The OpenFOAM variant called NavyFOAM shows high parallel efficiency during surface combatant wave-body interaction simulations and supports adaptation to classified hull designs [49]. HPC-based CFD serves as a tool for hull-propeller-rudder optimization which reduces cavitation and acoustic signatures in submarines although fully open submarine design data are limited [58, 85]. The ZNSFLOW technique utilizes domain decomposition to achieve zonal Navier–Stokes solutions as documented in CHSSI projects [32], yet literature references to external submarine flows do not disclose geometry and HPC scaling specifics. The development of HPC codes for wave energy converters designed to operate in large wave basins allows for the adaptation of their free-surface solvers to evaluate landing craft and advanced hull designs within realistic sea conditions which demonstrates the utility of open-source frameworks originally intended for renewable energy applications in naval defense operations [41].

### 6.2.1 Representative Table: Naval and Submarine Hydrodynamics

| Reference | HPC Approach | Open-Source Framework | Computing Resources | Mesh Scale | Turbulence Model | Key Findings |
|---|---|---|---|---|---|---|
| [49] | MPI-based domain decomp. | NavyFOAM (OpenFOAM fork) | Not specified | 3-13 million elements | RANS k-ω SST | Efficient free-surface simulation for ship hydrodynamics; near-linear scalability; validated wave-body interactions. |
| [58] | Parallel HPC solvers | Not specified | Not specified | 5.5 million cells | RANS/LES/DNS | Discusses HPC for marine industry, inc. propulsion and stealth aspects; partial references to submarine flow modeling. |
| [32] | Zonal domain approach | CHSSI HPC code (part open) | Defense HPC centers | Not specified | Not specified | Showcases HPC scaling for zonal solver in naval contexts; highlights large performance gains from HPC |
| | | | | | | modernization. |
| [41] | HPC wave modeling | OpenFOAM | Not specified | 14 million cells | Various models | Focus on wave energy converters but methods apply to wave-structure interactions for maritime defense scenarios. |

## 6.3 Space and Missile Defense

The simulation of space and missile defense systems requires high-performance computing-based computational fluid dynamics to model extreme Mach numbers and high-temperature gas dynamics alongside chemical nonequilibrium flows in scenarios that include reentry vehicles and ballistic interceptors which experience strong shocks and ablation effects with potential plasma phenomena [31]. Most advanced simulations stay classified but public studies provide some understanding of HPC-based frameworks and methods [74]. [18] open analysis of lateral jet interaction in the THAAD interceptor stands out as it demonstrates near-linear scaling across hundreds of processors with HPC-based RANS codes possibly linked to CHSSI CFD-7 or NPARC Alliance Flow Simulation System which reduced simulation times substantially. The NASA CFD Vision 2030 report emphasizes high-performance computing's benefits for chemically reactive hypersonic flows by mentioning adaptable defense agency codes FUN3D and DPLR [16, 75]. Scramjets also fall within HPC's purview: The supersonic combustion environment of scramjets requires dense grids and advanced turbulence/combustion models because of shock trains and stiff chemical reactions which need parallel implicit solvers [31]. The fluid simulation deep learning accelerator CFDNet shows potential for real-time design iteration due to its capability to decrease run times [63]. Numerical instabilities like shock instabilities and overshoots near steep gradients require the implementation of flux-splitting schemes along with limiters and advanced Riemann solvers which create extra complications for parallel domain interfaces when operating at large HPC scales [61].

## 6.4 Directed Energy and Other Emerging Technologies

High-energy lasers (HEL), railguns, and magnetohydrodynamic propulsion systems require high-performance computing fluid and thermal analysis despite the availability of limited open-source research information. The available public datasets occasionally consist of partial HPC-based MHD simulations and plasmadynamic models. According to [64], helicopters with laser systems may be tested for downwash and beam propagation effects but integrated laser-fluid coupling knowledge stays mostly restricted [64]. The development of HPC architectures for MHD propulsion requires solving fluid-electromagnetics equations which [11] anticipated would involve parallel codes combining extended Navier–Stokes with Maxwell's equations. Explicit open versions of multi-physics frameworks have limited documentation yet partial or developing modules likely exist building on OpenFOAM extensions [77]. The movement towards exascale computing in HPC prompts researchers to expect increased open-source or semi-open projects that model fluid dynamics in directed energy weapons and advanced propulsion technologies with data-driven methods speeding up parametric sweeps [63]. The classification barrier limits public-facing studies to presenting methodologies and partial code references instead of complete software implementations.

## 6.5 Security & Intelligence Applications

High-performance computing-enhanced computational fluid dynamics (HPC-accelerated CFD) supports homeland security and intelligence operations by modeling hazardous agent dispersal and performing flow analysis in urban environments to manage crowds and microclimates alongside protecting critical infrastructure including industrial accident assessments near nuclear facilities and refineries. The 2007 study by Patnaik and Boris demonstrated how HPC systems can be used for CBR defense through rapid simulations that assist first responders although their work remains not entirely open-source yet reflects the general trend of using HPC frameworks for immediate decision-making [64]. Through their 2009 research Senocak, Thibault, and Caylor verified GPU computing's acceleration capabilities by deploying NVIDIA clusters for nearly real-time CFD analysis of urban settings using RANS solvers while proposing open-source tool integration such as OpenFOAM [72]. High-performance computing (HPC) enabled computational fluid dynamics (CFD) supports the creation of nonlethal deterrents like high-intensity acoustic devices through fluid-structural interaction models which follow open research principles even when the devices remain proprietary. Environmental monitoring applications that predict toxin spread around critical infrastructure take advantage of large-scale unstructured meshing and parallel solvers which enable quick simulation of complicated urban flow fields according to [63]. The principal requirement for security-focused CFD applications involves delivering actionable insights within minutes or hours instead of days to meet the demands of dynamic operational situations [75]. Recent high-performance computing studies demonstrate the potential of machine learning surrogates and real-time sensor data integration to close this gap thus creating new routes to agile situational awareness in homeland security applications [64].

## 6.6 Cross-Domain Observations and Synthesis

The defense-oriented CFD research in airborne, naval, space, and homeland security applications shows several HPC techniques functioning as common components. MPI-based parallel decomposition stands out as a vital method for dividing extensive unstructured grids used in simulations from supersonic missile aerodynamics to submarine hull hydrodynamics. Open-source software like OpenFOAM and its extensions NavyFOAM and solids4Foam dominates defense-related open literature research which points toward a collective movement for using open frameworks in defense applications [49]. Hybrid turbulence models which integrate Large-Eddy Simulation (LES) with Reynolds-Averaged Navier–Stokes (RANS) methods are commonly used to analyze complex separated flow scenarios which simpler methods fail to detect. The review demonstrates that classification barriers persist in cutting-edge fields such as hypersonics and stealth technology where essential information on geometry, boundary conditions, and complete performance data continues to be limited or completely withheld. Despite restricted replicability older releases and partially declassified documents deliver sufficient details about numerical methods and HPC strategies that support knowledge transfer across different domains [19]. Advancing HPC-driven CFD in defense applications demands a team effort from fluid dynamicists alongside HPC software engineers and defense experts while machine learning experts are becoming increasingly crucial [64]. The collaborative methodology not only speeds up innovation by eliminating redundant code through shared repositories but also boosts verification processes through independent result replication while building capacity by training personnel in modern HPC methods at government labs and academic institutions [51]. The field is being transformed by several key trends including the adoption of GPUs which helps decrease both simulation time and energy use [27, 72], efforts towards exascale capabilities guided by NASA's CFD Vision 2030 [16, 75], and the rising use of machine learning for real-time defense CFD applications through surrogate modeling strategies [63].

### 6.6.1 Concluding Remarks on the Literature Survey

The study highlights that open-source HPC-accelerated CFD maintains a strong presence in publicly accessible research despite facing classification restrictions in defense-related final applications. HPC scaling demonstrates strong performance when applied to geometry-dense fluid simulations on airborne platforms such as high-lift aircraft UAVs and supersonic missiles. The field of naval hydrodynamics takes advantage of high-performance computing free-surface solvers through specialized OpenFOAM forks. High-performance computing-based computational fluid dynamics is essential for handling hypersonic or reentry flows in space and missile defense applications despite limited availability of open literature documentation. MHD-based propulsion and directed energy systems are still emerging technologies with potential applications in future high-performance computing developments. Near real-time hazard assessment is performed by security and intelligence communities utilizing HPC-driven dispersion modeling [2]. The research consistently points to advanced domain decomposition techniques alongside GPU acceleration for large-scale turbulence modeling while emphasizing the use of open-source frameworks to enable code reuse and promote collaborative development. Notably, references like [49], using HPC-based CFD capabilities to deliver near-linear scaling with precise outcomes have been established by [7, 18] and these qualities are essential for defense mission operations.

By analyzing these publications collectively, one can glean that HPC in defense context does not stand still: Exascale computing with $10^{18}$ FLOPS capabilities and machine learning surrogates will bring significant changes to fluid dynamics simulation methods in the next decade according to studies by [35, 75] with historical insights from [11]. The role of open-source CFD which was previously considered an accessory to security-focused research development has gained recognition as an efficient and flexible core support system because of governmental initiatives in HPC modernization that demand increased software transparency and compatibility. The reviewed references serve as a comprehensive guide for understanding technical achievements and persistent challenges while highlighting future directions in HPC-accelerated CFD applications for national security purposes [52].

# 7. Emerging Trends and Future Directions

## 7.1 Exascale Computing and Beyond

Exascale computing represents a major milestone since it achieves a performance level of at least $10^{18}$ floating-point operations per second which significantly surpasses early 21st century petascale systems. Through initiatives such as the U.S. Department of Energy's Exascale Computing Project (ECP) alongside Europe's EuroHPC initiative and similar projects in China and Japan exascale computing has become a fundamental resource for national security and scientific progress [62, 75]. Exascale HPC systems in the defense industry aim to overcome fluid dynamics problems which are fundamental to developing next-generation hypersonic vehicles as well as advanced propulsion systems and naval or aerospace design optimization. The unprecedented resolution and fidelity enabled by exascale platforms make it possible to model billions of mesh cells and capture transient multi-physics interactions which will revolutionize defense agencies' exploration of missile aerodynamics and stealth technologies as well as multiphase naval hydrodynamics [18, 49].

Exascale HPC promises capabilities that transcend mere computation speed. Direct numerical simulations (DNS) of high-Reynolds-number flows will become more feasible in ultra-high-fidelity simulations which will improve turbulence modeling for scramjet inlets and missile control surfaces as well as stealth configurations [14, 31]. Exascale systems will enhance CFD capabilities in defense applications to enable real-time responses for missile interception and urban hazard dispersion modeling [63, 72]. Increased computational power enables extensive design optimization and uncertainty analysis which allows researchers to examine large parameter spaces throughout vehicle shapes and modes of flight along with potential operational dangers [54]. Exascale computing advancements establish the foundation for defense digital twin development which uses virtual representations of complex systems through high-precision physics models and uncertainty analysis to decrease expensive physical tests.

To harness exascale CFD technology for defense applications requires addressing many significant hurdles. The transition to GPU-centric exascale systems like Frontier and Aurora requires extensive modification of traditional CPU-focused CFD software such as OpenFOAM and SU2 to maximize the benefits of heterogeneous computing environments [49, 51]. The necessity for fault tolerance becomes more challenging as exascale systems grow larger which raises the possibility of hardware failures demanding strong checkpoint-restart mechanisms and resilience strategies [11, 37]. Memory hierarchy complexities and communication bottlenecks have led researchers to adopt communication-avoiding algorithms alongside hybrid parallelism strategies to reduce data movement while increasing throughput according to [61]. Exascale computing advances beyond solver optimization to develop integrated multi-physics simulations that couple fluid solvers with structural, electromagnetic and thermal models through tightly organized workflows. The development of exascale-ready simulation platforms requires continuous partnerships between HPC centers, CFD developers, and defense engineers to maximize benefits for defense applications including hypersonic weapons and stealth aircraft.

## 7.2 AI, ML, and Data-Driven Methods

The fusion of high-performance computing (HPC) with artificial intelligence (AI) through machine learning (ML) stands as a pivotal development in computational fluid dynamics (CFD) for defense applications. Developers and engineers utilize surrogate models and reduced-order models (ROMs) to emulate high-fidelity simulations while greatly reducing the computational burden according to research by [15, 63]. Neural networks and regression-based surrogate models trained on HPC simulation data enable quick predictions of flow fields and aerodynamic performance across various geometries and operating conditions. The ability to deliver real-time decision support for missions and optimize designs within extensive parametric spaces proves essential for quickly evaluating small design adjustments. Recent advances involve physics-informed neural networks (PINNs) that integrate physical laws into neural network loss functions to guide outputs through partial differential equations as well as projection-based techniques such as proper orthogonal decomposition (POD) which transform CFD data into efficient surrogate models enabling quick iterative testing [54].

This synergistic approach primarily aims to improve turbulence modeling which remains a persistent challenge in defense-focused computational fluid dynamics. Researchers apply machine learning techniques to improve and transform turbulence closure models used in Reynolds-Averaged Navier–Stokes (RANS) and Large-Eddy Simulation (LES) frameworks [30]. Data-driven RANS corrections utilize high-fidelity DNS or LES data to improve turbulence viscosity models and stress-strain relationships while overcoming limitations found in standard k-ω and k-ε models particularly in separated flows and shock-boundary layer interactions [7, 20]. Neural network-based subgrid-scale models which have been trained using high-performance computing resolved datasets are now incorporated into LES solvers to incorporate localized physics into under-resolved turbulent scales [63]. The creation and confirmation of these models demand extensive HPC resources to achieve both numerical stability and accuracy for defense-relevant flow situations ranging from transonic through supersonic to hypersonic conditions.

Real-time or near real-time simulations in critical defense environments become possible through the collaboration of HPC and AI technologies. The integration of HPC-trained surrogate models with real-time sensor data leads to the development of digital twin frameworks that act as dynamic virtual copies of aircraft or missile systems providing immediate flow behavior predictions under changing operational scenarios [63, 72]. Adaptive CFD simulations integrate experimental or in-flight sensor data through AI-driven data assimilation workflows to dynamically adjust boundary conditions or turbulence models and enhance fidelity for advanced applications such as active flow control used in stealth optimization [24]. However, challenges persist. Access to comprehensive training data is limited due to the classified status of defense datasets which restricts the generalizability of surrogates in full-scale operational conditions [18]. Surrogate models struggle to predict scenarios outside their training distribution including uncommon hypersonic flow events or chemically reactive shock layers [58]. To reduce these risks organizations must keep funding uncertainty quantification efforts and regularly update training with new HPC datasets. When exascale computing reaches maturity phase AI integration within CFD workflows will become essential for fast data-driven defense simulation methods [11].

## 7.3 Quantum Computing Prospects

Quantum computing attracts a lot of interest because it could achieve exponential performance gains in specific algorithmic fields such as cryptographic factorization and unstructured search but its use in computational fluid dynamics (CFD) remains mostly theoretical. CFD's fundamental challenges originate from its grounding in continuous partial differential equations (PDEs) solutions and extensive iterative solvers which demand persistent floating-point precision beyond current quantum systems' capabilities [14]. Research into quantum linear algebra operations indicates potential future benefits for CFD problem-solving through quantum-accelerated matrix inversion and eigenvalue estimation methods [11]. Quantum computing is expected to gain initial success in defense applications through specialized wave-based simulations like radar cross-section modeling and quantum plasma interactions that benefit from quantum computing's natural ability to manage wave functions [68]. Quantum solutions for classical continuum flow problems defined by Navier–Stokes equations have not yet reached practical application levels because current quantum PDE solvers can only interpret extremely simplified models with toy geometries and ideal boundary conditions.

The most feasible near-term advancement involves hybrid quantum-classical architectures which have classical HPC platforms processing the bulk of CFD tasks while quantum processors enhance performance for selected computational bottlenecks including parts of linear solvers as well as optimization routines and uncertainty quantification tasks [75]. By hybridizing classical CFD solvers with quantum co-processors defense systems could transfer particular computational kernels to quantum processors which would result in improved performance for specified subroutine tasks. Current quantum hardware limitations including limited qubit availability, unstable gate operations, and excessive classical-system communication overhead prevent practical implementation. From a defense perspective, critical challenges persist: Existing quantum technology does not provide enough qubits to tackle the billions of degrees of freedom found in high-fidelity CFD simulations, and proper error correction systems are still many years to decades from development [65]. The defense community watches quantum computing progress for possible HPC applications like cryptography or radar modeling but expects quantum integration into operational CFD systems to emerge only after exascale computing becomes fully operational.

## 7.4 Interdisciplinary Collaborations

The complexity of modern military systems requires interdisciplinary collaboration as HPC-driven defense research now transcends fluid dynamics alone. High-fidelity CFD analysis combined with structural analysis remains essential for aerospace and naval applications to understand how aerodynamic or hydrodynamic forces interact with mechanical responses in elements like wings and control surfaces along with hulls and missile airframes [24]. High-performance computing allows closely integrated fluid-structure interaction (FSI) simulations that require dynamic boundary pressure and displacement exchanges between CFD software like OpenFOAM or SU2 and finite element structural analysis tools at each time step to maintain synchronized load and deformation patterns between domains [58]. The ability to accurately simulate rapid dynamic events like supersonic control surface movements and store ejection processes becomes essential in defense applications where aerodynamic precision and structural integrity must be maintained [21]. High Performance Computing systems improve the integration of nonlinear material models like temperature-sensitive composites and shape-memory alloys in UAVs and stealth platforms by enabling the simultaneous solution of coupled aero-thermal-structural phenomena under extreme conditions [31]. High-performance computing (HPC) facilitates multi-physics workflows that combine aerodynamics with electromagnetic simulations to improve vehicle shapes for enhanced aerodynamic efficiency as well as minimized radar cross-section (RCS). Defense engineers can balance stealth objectives with aerodynamic performance through simultaneous execution of CFD and EM solvers inside HPC systems [54, 75]. The same co-simulation methodologies apply to electromagnetic compatibility (EMC) analysis which uses CFD-modeled cooling airflow and EM solvers to protect electronic subsystems from resonance or interference.

HPC systems provide interdisciplinary capabilities which include environmental modeling along with the development of new defense technologies. Operational planning now integrates CFD models of localized airflows including crosswinds on aircraft carriers and runway vortex behavior with mesoscale weather forecasts generated by HPC-based climate models to boost readiness in weather-sensitive theaters [64, 72]. High-performance computing drives defense innovation through multi-physics simulations that combine fluid dynamics with acoustics, thermostructural mechanics, and electromagnetics. The development of hypersonic vehicles demonstrates this integration through the coupling of CFD solvers with thermal protection system ablation models and both reentry plasma dynamics and material fatigue simulations under extreme heat fluxes. Naval hydrodynamic simulation technologies advance as they connect with acoustic propagation software to assess submarine detectability by analyzing flow-induced noise alongside sonar scattering features for stealth optimization [45, 49]. Directed energy applications, such as high-energy laser systems, further illustrate this trend: High-performance computing based computational fluid dynamics (CFD) enables modeling of laser-induced shock waves and turbulent plumes while high-performance computing optical solvers predict beam distortion or attenuation. To coordinate complex models that span multiple domains requires scalable HPC frameworks that can handle various time-step constraints and discretization methods [51]. The synergy of multi-physics fields through High-Performance Computing architectures leads to advanced digital twins which can simulate defense-related phenomena with unmatched realism and operational efficiency despite logistical challenges.

## 7.5 Conclusion on Future Directions

The development path of HPC-accelerated CFD within defense sectors clearly leads to an exascale computing future with AI-driven processes and comprehensive multi-physics systems integration. Moving from petascale systems to exascale systems and further presents numerous obstacles. The progression to exascale computing in defense agencies involves significant financial and organizational challenges including the update of legacy CFD software and training engineers to handle diverse computing architectures and complex memory management systems [16, 75]. Security restrictions that limit international cooperation and challenge the open-source principles which drove CFD innovation complicate these technical challenges [59]. The Department of Defense along with partner organizations now use digital engineering principles which make high-performance computing driven computational fluid dynamics the essential component of virtual prototyping processes to quickly explore enormous design possibilities before building physical models. Operational domains are experiencing transformation through the combination of exascale capacity and AI-powered surrogates as partial real-time CFD simulations paired with live sensor data promise to strengthen situational awareness for applications such as missile defense, urban security, and CBR hazard response [63, 71, 72]. Defense HPC workflows have transitioned from offline analyses to become interactive decision-support tools that guide operators throughout essential mission phases.

The future direction of defense CFD encompasses both interdisciplinary collaboration and extensive computational requirements. Exascale and post-exascale challenges are being addressed by multi-sector ecosystems that unite national laboratories with private sector HPC vendors, academic institutions and open-source communities through tightly integrated collaborations [32, 51]. Despite quantum computing and neuromorphic hardware being distant technologies, classical HPC systems with AI accelerator enhancements will maintain their dominance over CFD workflows in the foreseeable future. Defense stakeholders maintain vigilant oversight over these new technologies because they understand their ability to transform traditional modeling approaches and merge simulation activities with real-world operations. The modern defense strategy relies heavily on HPC-accelerated CFD because it allows engineers and decision-makers to simulate complex multi-physics environments with unmatched fidelity and speed. Defense agencies face a dual challenge as they implement digital twins and enhance hypersonic designs alongside electromagnetic and structural integration in unified HPC systems: they must utilize massive computational resources while simultaneously promoting secure and collaborative innovation that addresses the changing needs of international security.

## 8. Conclusions

Modern defense and national security innovation now depend on High-Performance Computing (HPC) and Computational Fluid Dynamics (CFD) which allow experts to model sophisticated aerodynamic, hydrodynamic, and multi-physics phenomena with exceptional accuracy. The examination validates that high-performance computing powered computational fluid dynamics remains essential across applications including missile defense systems and naval hydrodynamics along with urban security measures. The defense sector together with research organizations now depend on extensive parallel architectures and sophisticated turbulence modeling along with high-resolution simulations to reduce design timelines and boost mission dependability [18, 75]. High-performance computing-enabled CFD has revolutionized aerodynamic prediction in aerospace technology by solving complex flow issues including vortex breakdown and high-angle-of-attack instabilities which affect next-generation aircraft and interceptors [7, 24]. The Navy uses free-surface flow modeling and cavitation control with NavyFOAM for naval applications while submarine stealth optimization relies on HPC-driven multiphase simulations to reduce acoustic and visual signatures. High-performance computing (HPC) remains indispensable for advanced missile defense systems and propulsion concepts because it enables the resolution of complex multi-physics scenarios that include shock-wave interactions and chemically reacting flows [31, 62]. Homeland security initiatives use GPU-accelerated computational fluid dynamics to model urban dispersion of chemical, biological, and radiological threats in real time which supports strategic improvements and infrastructure durability [64, 72]. HPC delivers simulations which maintain necessary scale and timely results that drive defense agencies in their development and implementation of advanced technology solutions.

The growing importance of open-source contributions stands out within this highly specialized field. Although classification constraints exist, an active network of open frameworks including OpenFOAM and SU2 alongside domain-specific versions promotes HPC-based CFD advancements for defense applications [7]. The modular design of OpenFOAM facilitates user-defined solver creation for high-lift aerodynamics and hypersonic vehicle design applications and NavyFOAM demonstrates how open frameworks can adapt to naval hydrodynamics needs [49]. The SU2 framework demonstrates successful academic-defense collaboration through its focus on aerodynamic optimization [54]. The historical development of legacy HPC codes from CHSSI projects such as NPARC and ZNSFLOW serves as a foundational platform for current parallelization techniques and solver strategies. Modern progressions demonstrate faster workflows for GPU-based computations and artificial intelligence systems within urban CFD simulations and surrogate model training pipelines [63, 72]. The fundamental algorithms, HPC scaling techniques, and verification methodologies are becoming more publicly available through open-source channels while defense-specific datasets and classified geometries stay under strict control. Defense organizations can leverage this collective knowledge repository to accelerate solver development and stimulate cross-sector innovation by constructing secure solutions on open HPC platforms while ensuring strict security protocol adherence.

Open-source high-performance computing frameworks continue to gain traction in defense CFD applications but face significant obstacles when applied to environments with high classification levels. Export regulations under ITAR and EAR create strong limitations that prevent sharing of algorithms or datasets which might improve adversarial capacities according to [59]. Open-source development promotes creative advances but faces fundamental limitations due to potential exposure of protected geometrical designs and performance metrics in public code bases. Critical defense modules such as proprietary turbulence closures and stealth-specific meshing strategies stay confined within closed-source defense contractor systems which generates integration difficulties with public CFD platforms [51]. The air-gapped configuration of defense HPC clusters restricts rapid adoption of updates or community patches which makes code modernization progress slower than in the open-source ecosystem [46, 83]. Open-source CFD provides essential testing grounds for scaling and methodological advancements yet remains inherently limited for direct application to next-generation classified systems [44]. Open literature mainly focuses on outdated legacy systems and scaled-down models which creates a disconnect between academic advancements and the latest developments in secure defense systems.

The analysis uncovers multiple essential research deficiencies within the open-source HPC CFD area that exist outside established security boundaries. The publicly available research lacks sufficient studies on high-Mach-number reentry flows and plasma-field interactions as well as chemically reacting hypersonic flows. Current HPC-based predictions at the outer limits of defense applications lack validation because substantial datasets for shock-dominated reentry aerothermodynamics and advanced scramjet combustor flows do not exist [75]. Open repositories inadequately cover multiphase flow applications involving cavitation and fluid-structure-acoustic coupling for submarine stealth since only a limited number of end-to-end simulations connect hydrodynamics with sonar scattering. Machine learning surrogates and data-driven turbulence models demonstrate significant potential yet remain largely experimental due to challenges in merging large HPC-generated datasets with limited flight or sea trial data [13, 63]. Furthermore, disparities in access to HPC infrastructure exacerbate this issue: Major contractors and national laboratories have access to dedicated supercomputing resources while smaller research groups encounter limitations in computing cycles and face costly resource expenses [51]. The current limitations highlight the importance of establishing broad multi-institutional partnerships and fostering better resource-sharing programs alongside creating open high-quality benchmark datasets to advance the defense HPC sector.

# 9. Recommendations

Defense organizations must implement disciplined software design and governance practices to effectively operate open-source HPC CFD frameworks within security-sensitive environments. Modular code structures offer an effective approach that lets a publicly accessible core solver enable community interaction while segregated branches maintain classified proprietary modules. Modular structures facilitate wide-ranging collaborative development of scalable HPC algorithms while protecting sensitive technologies like stealth shaping and proprietary turbulence models. Defense-associated HPC repositories require rigorous protocols that include mandatory signed commits, continuous integration pipelines and independent code audits to ensure security and trust [25]. Sanitized boundary data with abstracted geometries and generic flow conditions enables large-scale testing of open-source solvers while protecting sensitive system details. HPC-accelerated CFD development continues to meet security requirements while maintaining scientific transparency through a balanced approach between openness and discretion.

The development of tailored verification and reproducibility standards for HPC systems holds equal importance. High-stakes defense decisions rely on HPC-based CFD simulations which necessitates that current validation and verification (V&V) frameworks such as AIAA or NASA guidelines must adapt to address extreme-scale parallelism and complex multi-physics workflows [7, 24]. Solver robustness under exascale and hybrid CPU-GPU systems will benefit from the integration of performance scaling tests as well as dynamic load-balancing assessments and regression benchmarks according to studies by [12, 51]. Containerization strategies and HPC-specific deployment recipes can reinforce reproducibility by maintaining the portability of solver environments across various HPC infrastructures. Defense agencies need to promote initiatives for partial data release like sanitized geometries and benchmark cases to build collaborative relationships between academic institutions and industry as well as government labs. Open HPC competitions and cross-institutional test facilities hold potential to drive innovation in specialized areas including high-lift aerodynamics and naval stealth modeling while respecting national security limitations.

The field of open-source HPC CFD will need specific research initiatives to maintain its relevance within defense applications going forward. Defense applications like missile plume interactions and submarine wake flows require curated benchmark libraries similar to NASA's CRM. To prepare for exascale computing it is essential to prioritize collaborative research on GPU kernelization alongside fault tolerance and data compression to safeguard codes for upcoming advanced architectures [16, 63]. Open-source initiatives should simultaneously advance machine learning integrated turbulence closures and surrogate models that have undergone validation through wind tunnel experiments or restricted-access flight data studies [7, 87]. The development of hybrid multi-physics approaches which integrate fluid dynamics with structural analysis and thermal and electromagnetic simulations will enhance the capabilities of open-source defense models [31]. The defense community will reach the full capabilities of open-source HPC-based CFD for next-generation military technology by investing strategically in interdisciplinary scalable initiatives combined with transparent international collaboration under well-defined security protocols.